\documentclass[12pt,preprint]{aastex}

\def\spose#1{\hbox to 0pt{#1\hss}}

\def\etal{{\em et al.}}

\def\arcsec{\hbox{$^{\prime\prime}$}}

\def\lsim{\mathrel{\hbox{\rlap{\lower.55ex \hbox {$\sim$}}\kern-.0em\raise.4ex \hbox{$<$}}}} 
\def\gsim{\mathrel{\hbox{\rlap{\lower.55ex \hbox {$\sim$}}\kern-.0em\raise.4ex \hbox{$>$}}}}

\def\grb{GRB\thinspace{980703}}
\def\ts{\thinspace}

\def\ie{{\em i.e.,}}

\begin{document}

\title{The Broadband Afterglow of GRB\thinspace{980703}}

\author{D. A. Frail\altaffilmark{1,2},
S. A. Yost\altaffilmark{1},
E.    Berger\altaffilmark{1},
F. A. Harrison\altaffilmark{1},
R.    Sari\altaffilmark{3},
S. R. Kulkarni\altaffilmark{1},
G. B. Taylor\altaffilmark{2}, 
J. S. Bloom\altaffilmark{1},
D. W. Fox\altaffilmark{1}
G. H. Moriarty-Schieven\altaffilmark{4}
P. A. Price\altaffilmark{1,5},
}

\begin{abstract}
  We present radio observations of the afterglow of the bright
  $\gamma$-ray burst \grb\ made between one day and one year after the
  burst. These data are combined with published late-time radio
  measurements and existing optical, near-infrared (NIR) and X-ray
  observations to create a comprehensive broadband dataset for
  modeling the physical parameters of the outflow. While a
  wind-stratified medium cannot be ruled out statistically, it
  requires a high fraction of the shock energy in the electrons, and
  so is not favored on theoretical grounds. Instead, the data are
  consistent with a fireball model in which the ejecta are collimated
  and expanding into a constant density medium. The radio data cannot
  be fit with an isotropic shock but instead require a jet break at
  $\approx$ 3.5 days, not seen at optical wavelengths due to the
  presence of a a bright host galaxy.  The addition of the full radio
  dataset constrains the self-absorption frequency, giving an estimate
  of the circumburst density of $n \approx 30$ cm$^{-3}$, a value
  which differs substantially from previous estimates.  This result is
  consistent with the growing number of GRB afterglows for which
  broadband modeling yields $n \simeq 0.1-100$ cm$^{-3}$, with a
  typical value $\sim$ 10 cm$^{-3}$.
\end{abstract}

\slugcomment{Submitted to The Astrophysical Journal}

\altaffiltext{1}{Division of Physics, Mathematics and Astronomy,
  105-24, California Institute of Technology, Pasadena, CA 91125}

\altaffiltext{2}{National Radio Astronomy Observatory, P.O. BOX `O', Socorro,
NM 87801} 

\altaffiltext{3}{Theoretical Astrophysics 130-33, California Institute
  of Technology, Pasadena, CA 91125}

\altaffiltext{4}{National Research Council of Canada, Joint Astronomy Centre, 
660 N. A'ohoku Place Hilo, HI 96720}

\altaffiltext{5}{Research School of Astronomy and Astrophysics, Mount 
Stromlo Observatory, via Cotter Rd., Watson Creek 2611, Australia}


\newpage
\section{Introduction}

Astronomers have monitored the afterglows of GRBs with considerable
enthusiasm across the electromagnetic spectrum. The primary motivation
in using these measurements is to infer the fundamental parameters of
the explosion: the total energy release, the geometry of the explosion
and the density distribution of ambient gas \cite{wg99, cl99, hys+01,
  pk01}.

The gamma-ray burst (GRB) of 1998 July 3.18 UT triggered the BATSE
detectors on board the {\it Compton Gamma-Ray Observatory} \cite{k+98}
and its afterglow was detected by the All Sky Monitor on the the {\it
  Rossi X-ray Timing Explorer} (Levine Morgan \& Muno
1998)\nocite{lmm98}. Followup observations of the X-ray afterglow were
obtained with the Narrow Field Instruments (NFI) on the {\it BeppoSAX}
satellite (see Vreeswijk et al.~1999 for a summary of NFI
observations). Radio observations of this field with the Very Large
Array (VLA) began 1.2 days after the burst and identified a radio
source within the {\it BeppoSAX} NFI error circle.  Coincident with
this we discovered a fading optical source and suggested that the
source was the radio and optical afterglow of \grb\ (Frail et
al.~1998)\nocite{fhb+98}. Zapatero-Osorio et al.~(1998)\nocite{zgo+98}
also reported the same fading optical source, while Djorgovski et al.
(1998)\nocite{dkg+98} discovered the host galaxy and measured its
redshift $z=0.966$. See Bloom et al.~(1998)\nocite{bfk+98} for a
summary of the early radio, optical and NIR measurements.

Unfortunately, the host galaxy of \grb\ is very bright -- $R\sim 22.6$
magnitude \cite{dkb+98b} and so while this has led to a number of
interesting results regarding the physical properties of GRB host
galaxies ({\em i.e.}, Holland et al.~2001\nocite{hfh+01}; Sokolov et
al.~2001\nocite{sfc+01}; Berger, Kulkarni \& Frail 2001\nocite{bkf01};
Chary, Becklin \& Armus 2002\nocite{cba02}), it has also meant that
the optical and NIR afterglow could be tracked for only a few days
before it faded below the light from the host galaxy. For this reason,
the temporal decay of the optical/NIR afterglow is poorly constrained
with $\alpha$ ranging from $-1.17\pm 0.25$ to $-1.61\pm 0.12$ (Bloom
et al.~1998; Castro-Tirado et al.~1999; Vreeswijk et al.~1999; Holland
et al.~2001); here flux at time $t$, $f(t)\propto t^{\alpha}$.
Furthermore, the host galaxy appears to be undergoing vigorous star
formation and consequently has a large amount of dust and gas
\cite{dkb+98b, sfc+01}.  Not surprisingly, the optical/NIR spectrum of
the afterglow (usually characterized by a power law, $f(\nu)\propto
\nu^{\beta}$, appears to be significantly affected by extinction
within the host galaxy.  The low precision with which $\alpha$ and
$\beta$ were measured preclude constraining the fundamental explosion
parameters with any reasonably precision (Bloom et al.~1998; Vreeswijk
et al.~1999).

Fortunately, the radio afterglow of \grb\ was quite bright and as a
result we were able to mount an ambitious monitoring program at the
Very Large Array (VLA). Here we present our final results on the
centimeter radio light curves of \grb\ and then proceed to interpret
the observations in the framework of afterglow models.  The primary
advantage of the radio measurements is the immunity of the radio
emission from the two effects discussed above (bright host and
extinction).  As a result, by combining the X-ray, optical/NIR and
radio data together in a single broadband dataset, we are able to
infer the physical parameters for the afterglow from \grb\ with
moderate precision.

\section{Observations and Data Reduction}\label{sec:obs}

The details on the initial discovery of the radio afterglow from \grb\ 
are given in Bloom \etal\ (1998). The late-time data ($\Delta{t}>300$
days) for this burst have been published by Berger, Kulkarni \& Frail
(2001). Below we describe the VLA monitoring program and observations
at other radio facilities.

\noindent{\it Very Large Array (VLA).}
VLA\footnotemark\footnotetext{The NRAO is a facility of the National
  Science Foundation operated under cooperative agreement by
  Associated Universities, Inc. NRAO operates the VLA and the VLBA.}
observations and data reduction were carried out following standard
practice. To maximize sensitivity the full VLA continuum bandwidth
(100 MHz) was recorded in two 50 MHz bands, each with both hands of
circular polarization. The flux density scale was tied to
3C\thinspace{48} (J0137+331) and frequent observations (every 2-5
minutes) were made of the phase calibrators J2346+095 (at 4.86 and
8.46 GHz) and J2330+110 (at 1.43 GHz). A log of the observations,
giving the measured fluxes at 1.43, 4.86 and 8.46 GHz, can be found in
Table~\ref{tab:radio}.

One VLA observation was made at 15 GHz on 1998 July 17.56 UT employing
the same methodology. No source was detected at 15 GHz above a
3$\sigma$ limit of 1.0 mJy.

\noindent{\it James Clerk Maxwell Telescope (JCMT).}
An observation was made on 1998 July 10.53 UT using the SCUBA array on
JCMT\footnotemark\footnotetext{The James Clerk Maxwell Telescope is
  operated by The Joint Astronomy Centre on behalf of the Particle
  Physics and Astronomy Research Council of the United Kingdom, the
  Netherlands Organization for Scientific Research, and the National
  Research Council of Canada.} at 220 GHz. The planet Uranus was used
as a primary flux calibrator.  The data were reduced in the standard
method (\ie\ corrected for atmospheric opacity which is estimated by
extrapolating from a skydip made at 225 GHz by a radiometer operated
by the Caltech Submillimeter Observatory) and converted to mJy based
on the primary flux calibrator.  The pointing was checked immediately
before and after the observations on a nearby blazar and was found to
vary by less than $\sim$2\arcsec. Despite excellent photometric
conditions, no 220 GHz source was visible at the position of \grb\ 
above a 2$\sigma$ limit of 5.2 mJy. Similar upper limits from JCMT are
reported by Smith \etal\ (1999)\nocite{stv+99}.

\noindent{\it Very Long Baseline Array (VLBA).}
Very Long Baseline Interferometry (VLBI) observations were performed
on 1998 August 2 at 8.42 GHz, using the 10 element VLBA for 5.6 hours.
Both right and left circular polarizations were recorded using 2 bit
sampling across a bandwidth of 32 MHz.  The VLBA correlator produced
16 frequency channels across each 8 MHz IF during every 2 second
integration.  Amplitude calibration for each antenna was derived from
measurements of the antenna gain and system temperatures.  Global
fringe fitting was performed on the strong nearby calibrator
J2346+0930 and the resulting delays, rates and phases were transfered
to \grb\ before averaging in frequency or time.  The time for a
complete cycle on the phase calibrator and target source was 3
minutes.

The data for all sources were edited, averaged over 30 second
intervals, and then imaged using DIFMAP \cite{shep97}. We detected
\grb\ with the VLBA at a level of 0.58 $\pm$ 0.06 mJy, consistent with
VLA measurements at this same time. At the time of the VLBA
observation, we place a limit on the angular size of the radio
afterglow of \grb\ of $<$0.3 mas. We also derive a position of
$\alpha$(J2000)~=~23$^{\rm h}$59$^{\rm m}$06\rlap{.}{$^ {\rm s}$}6661,
$\delta$(J2000)~=~8$^\circ$35$^{\prime}$07\rlap{.}{$^{\prime\prime}$}0939
with an uncertainty of 0.0007 arcsec in each coordinate.

\section{Broadband Data}\label{sec:bbdata}

\begin{figure}
        \epsscale{0.85}
\plotone{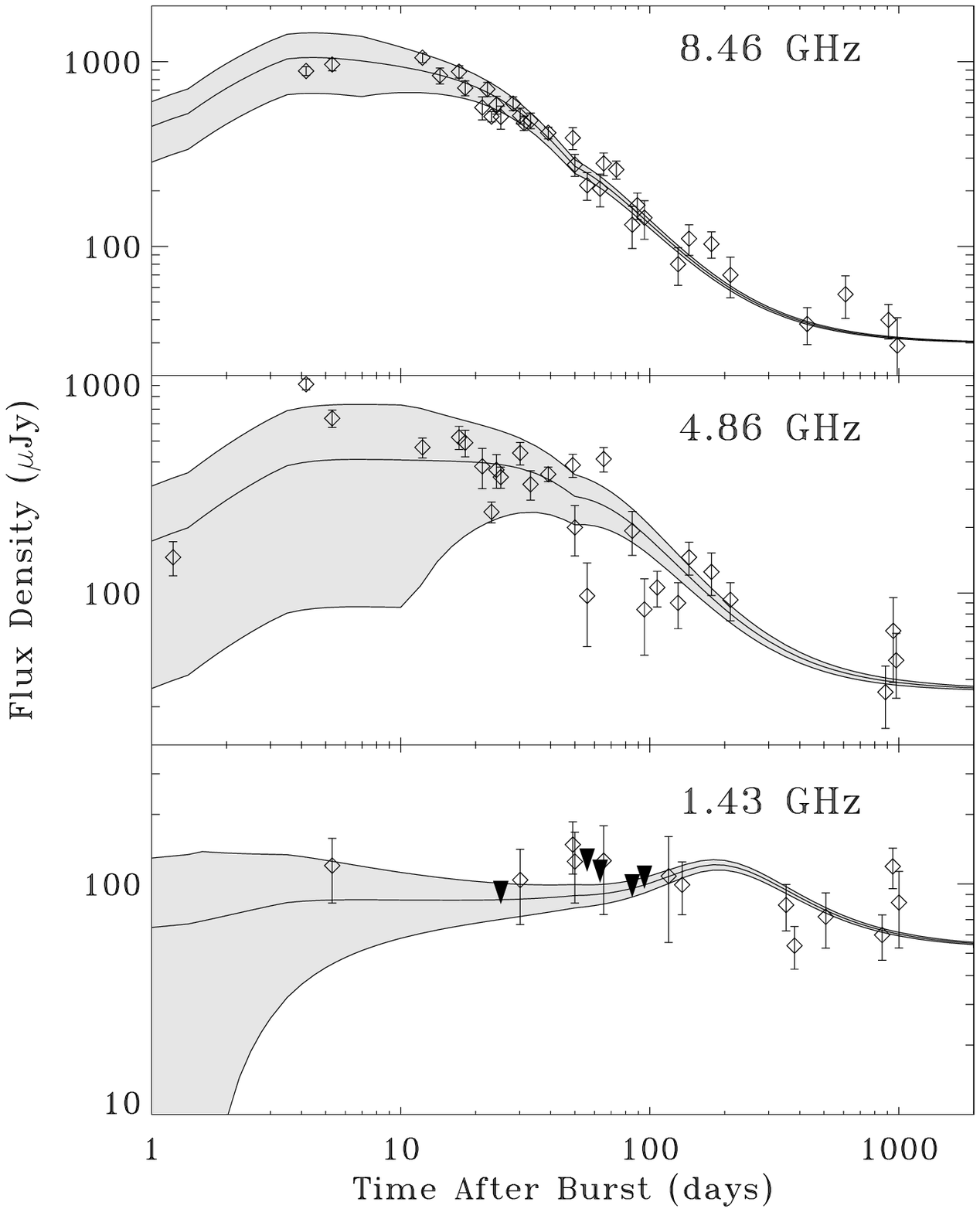}
\caption{Radio lightcurves of the \grb\ afterglow. The solid line is
  the best-fit model (see text for details), and the model lightcurves
  are plotted with their calculated 1-$\sigma$ scintillation
  envelopes. Upper limits (solid triangles) are plotted as the flux at
  the position of the afterglow plus two times the rms noise.
\label{fig:radio}}
\end{figure}

\begin{figure}
\plotone{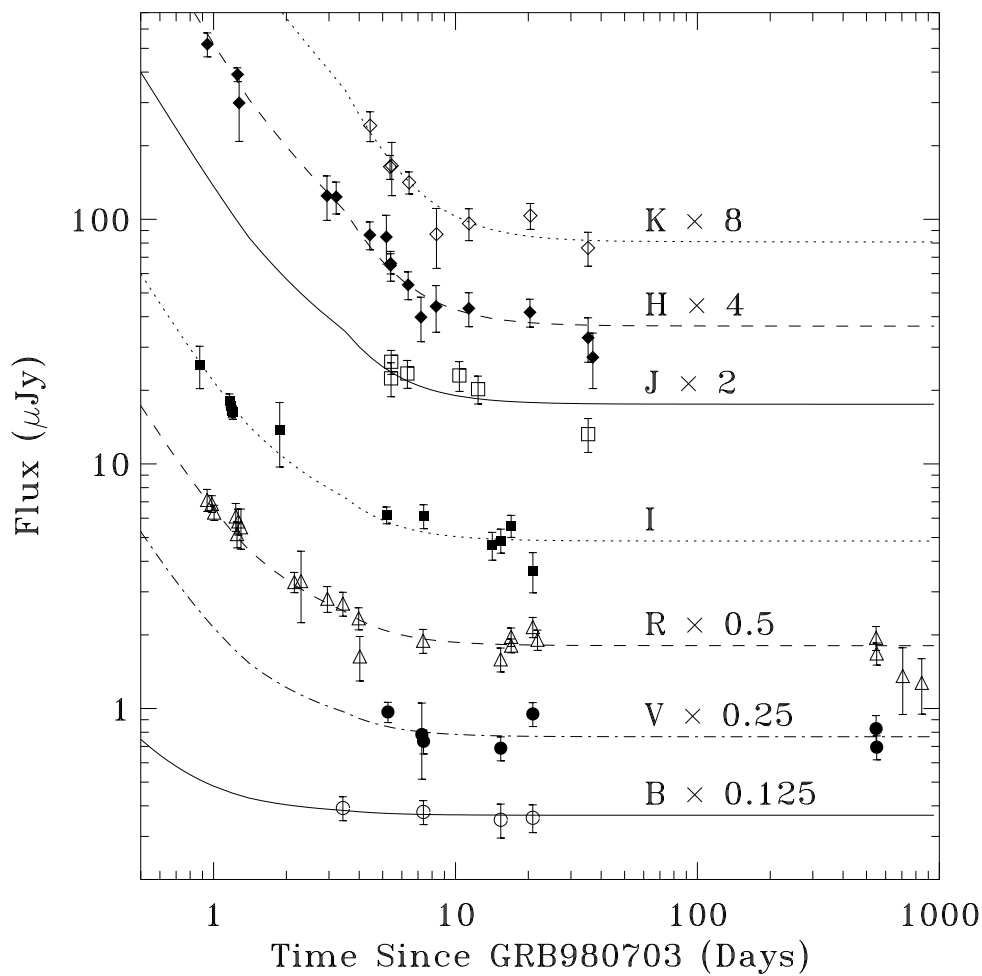}
\caption{Optical/NIR lightcurves of the \grb\ afterglow. The best-fit
  model is shown by the lines (see text for details). For ease of
  viewing the flux at each band has been multiplied by the factor
  given. The data are corrected for Galactic (but not host)
  extinction.
\label{fig:optical}}
\end{figure}

\begin{figure}
\plotone{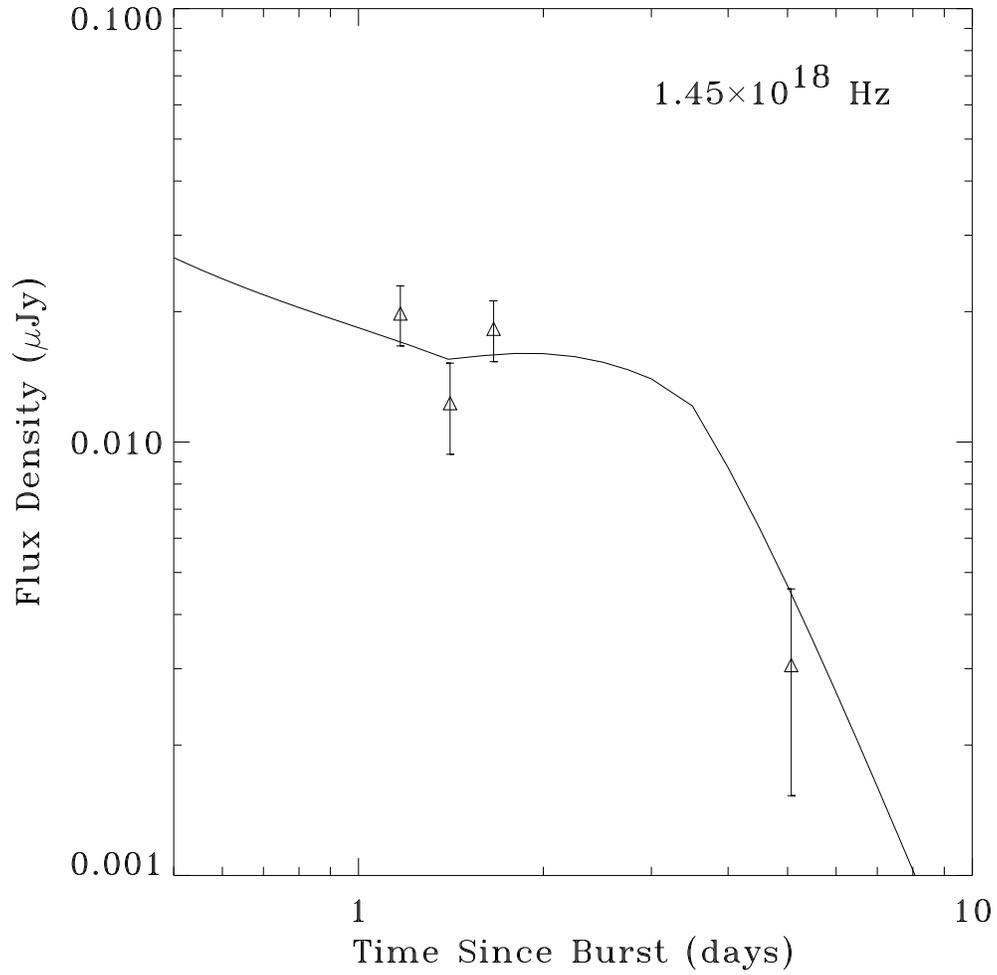}
\caption{The X-ray lightcurve of the \grb\ afterglow. 
  The best-fit model is shown with solid lightcurves (see text for
  details). The curvature seen in the model after the first day is the
  signature of a significant inverse Compton contribution to the X-ray
  afterglow flux at that time.
\label{fig:xray}}
\end{figure}

Before undertaking any detailed model fits it is worthwhile to review
the general characteristics of the entire broadband dataset for this
afterglow. In addition to the radio data summarized in \S\ref{sec:obs}
and Table~\ref{tab:radio}, there exists a large amount of published
data in the X-ray \cite{vgo+99}, optical/NIR
\cite{bfk+98,czg+99,vgo+99,hfh+01,sfc+01} and radio (Berger \etal\ 
2001)\nocite{bkf01} bands. Light curves of these data are plotted in
Figs.~\ref{fig:radio}-\ref{fig:xray}. The X-ray measurements were
converted to flux density with the spectrally-weighted factor (using
the observed photon index) that 1~Jy = 2.4$\times 10^{-11}$ erg
cm$^{-2}$ s$^{-1}$. We corrected the optical data for absorption in
our Galaxy (Schlegel, Finkbeiner \& Davis 1998)\nocite{sfd98} before
converting to flux densities using the factors in Bessell
(1979)\nocite{b79} for the optical and Bessell \& Brett
(1988)\nocite{bb88} for the near-IR bands. An additional 1\% error was
added in quadrature to all the measured flux densities to account for
any cross-calibration systematic uncertainties.

In Fig.~\ref{fig:radio} we display the radio light curves at the
frequencies of 1.43, 4.86 and 8.46 GHz. The 8.46 GHz light curve has a
well-defined peak above 1 mJy between 5 and 12 days after the burst,
followed by a power-law decay. As noted previously by Berger \etal\ 
(2001), the flux density at centimeter wavelengths undergoes a
flattening about 1 year after the burst which is attributed to
synchrotron emission from an underlying host galaxy.  After
subtracting this component (F$_{\rm host}$=39 $\mu$Jy) from the 8.46
GHz light curve we derive a temporal decay index
$\alpha_R=-1.05\pm0.03$ (where F$_R\propto t^{\alpha_R}$) between 12
and 1000 days after the burst.

The 4.86-GHz light curve shows a similar rise and a decay as that at
8.46 GHz. However, superimposed on this long-term secular behavior
there are significant changes in the flux density from one point to
the next.  These erratic fluctuations are not confined to day-to-day
variations but there is also evidence for short-term variability
(50\%) on timescales of a few hours. Narrow-band, short-timescale flux
variations are a hallmark of interstellar scattering (ISS)
\citep{goo97,fkn+97}. Although we make rough approximations for the
ISS-induced fluctuations in \S\ref{sec:bbmodel}, a more detailed
treatment of ISS for \grb\ is postponed for a later paper.

In contrast to the flux variations seen at 8.46 GHz and 4.86-GHz, the
1.43 GHz light curve is notable for its relative constancy. Most of
the emission at this frequency is dominated by the host galaxy with
F$_{\rm host}\sim$68 $\mu$Jy (Berger et al.~2001). After allowing for
some variation due to ISS, the peak flux of 0.15 mJy reached $\sim$50
days after the burst is well below the peak at 8.46 GHz ($\sim$1 mJy)
and at 4.86 GHz ($\sim$0.3 mJy).  This apparent drop in the peak flux
density with decreasing frequency (\ie\ ``peak flux cascade'') has
been noted for other well-studied bursts (Frail, Waxman \& Kulkarni
2000\nocite{fwk00}, Yost et al.~2002\nocite{yfh+02}) and poses an
important constraint on possible models (see \S\ref{sec:bbmodel}).

The optical/NIR data shown in Fig.~\ref{fig:optical} exhibits the
familiar power-law decay of the afterglow.  \grb\ occurred in a bright
GRB host galaxy \citep{dkb+98b} and so the optical/NIR afterglow could
only be followed for a few days before the host dominated the light
curve. The B, V, R, I, J, H, and K band light curves can be
characterized by a power-law afterglow component (in time and
frequency) plus a frequency-dependent host component. There is also a
small excess in the flux density between the R and K bands near day
20. As noted by Holland et al.~2001\nocite{hfh+01}, this could be due
to a supernova component in the late-time light curve but its
significance is not strong enough to warrant its inclusion in the
fitting.

A noise-weighted least squares fit of the form F$(\nu,t)$=\ F$_\circ
t^\alpha \nu^\beta$ + F$_{\rm host}(\nu)$ was carried out on the
entire optical/NIR data and yielded\footnotemark\footnotetext{The
  fitted F$_{\rm host}(\nu)$ are not given here since more accurate
  values are discussed in \S\ref{sec:bbmodel} and listed in
  Table~\ref{tab:model}} $\alpha_o=-1.67\pm0.08$ and
$\beta_o=-2.67\pm0.08$ with $\chi^2/dof$=64.7/66. The steep spectral
slope $\beta_o$ relative to the X-ray ($\beta_X=-1.51 \pm 0.32$) has
been noted before and attributed to dust extinction from the host
galaxy \citep{vgo+99}.  Our more accurate value of $\alpha_o$ is
consistent with earlier derivations \cite{bfk+98,czg+99,vgo+99}, but
it is considerably steeper than the radio ($\alpha_R=-1.05$) and X-ray
($\alpha_X<-0.91$) light curves in Figs.~\ref{fig:radio} and
\ref{fig:xray}.

\section{Broadband Modeling}\label{sec:bbmodel}

We interpret the observations summarized in \S\ref{sec:obs} and
\S\ref{sec:bbdata} within the framework of the standard relativistic
blast wave model (see M\'esz\'aros 2002\nocite{mes02} for a review).
In this model an impulsive release of energy from the GRB event drives
an ultra-relativistic outflow into the surrounding medium. Particle
acceleration occurring within this forward-propagating shock produces
the afterglow emission via synchrotron and/or the inverse Compton
processes.  Since the evolution of the blast wave is sensitive to the
energy and geometry of the explosion, as well as the density structure
of the circumburst medium, the modeling of the afterglow emission can
be used, in principle, to extract valuable information on GRB
progenitors and their environments, as well as details on the
microphysics of the shock (e.g. Panaitescu \& Kumar
2001a\nocite{pk01b}).

The particular approach we have taken to model broadband afterglow
emission has been described in some detail in two recent papers
\cite{hys+01, yfh+02}. In brief, we characterize the broadband
spectrum by several break frequencies, including both synchrotron and
inverse Compton components, one of which usually dominates depending
on the circumstances. The microphysics of the shock, such as the
electron energy index, $p$, the fraction of shock energy in electrons
$\epsilon_e$, and the fraction of shock energy in magnetic field
$\epsilon_B$ are taken to be invariant with time. The temporal
evolution of the break frequencies is governed by the energy of the
shock (which can be radiative), the geometry of the shock (which can
be isotropic or jet-like), and the density structure of the
surrounding medium (which can be constant or vary as the inverse
square of the radius). In addition to the basic physics, the model
also accounts for several complicating effects such as ISS at radio
wavelengths, dust extinction in the optical/NIR bands, and a possible
pan-chromatic contribution to the emission from a host galaxy.

The solution which best describes all the afterglow data for \grb\ is
a collimated outflow expanding into a constant density medium. Under
the heading ``ISM'' Table~\ref{tab:model} summarizes the best-fit
parameters which were derived using a least-squares approach. In
addition to the shock parameters $p$, $\epsilon_e$, and $\epsilon_B$,
the model solves for the jet opening angle $\theta_{jet}$, the
circumburst density $n$, the isotropic-equivalent fireball energy at
the time when the fireball evolution becomes largely adiabatic $E_{\rm
  iso}(t_{\nu_c = \nu_m})$, the restframe extinction A(V), and the
host flux density at several wavelengths. Perhaps the most striking
feature of this model is that it requires a jet break at $\sim 3.5$
days after the burst. The expected steepening of the optical/NIR
lightcurves at $t_{jet}$ is not obvious because of the brightness of
the host galaxy.  Although the steep value of $\alpha_o$ relative to
$\alpha_R$ and $\alpha_X$ is suggestive (see \S\ref{sec:bbdata} and
Holland et al.~2001), the case for a jet in \grb\ is based primarily
on the peak flux cascade observed at radio wavelengths (see
\S\ref{sec:bbdata} and Fig.~\ref{fig:radio}). It is this same behavior
that makes it impossible to model the afterglow of \grb\ as an
adiabatic expansion of an isotropic shock. In general, since radio
afterglows exhibit a different observational signature than either
that of optical or X-ray afterglows, they have proven useful in
revealing other cases of ``hidden jets'' \cite{bdf+01}.

Now that the true geometry is known ({\em i.e.,} $\theta_{jet}\sim
13^\circ$), the energy released in the GRB phase $E_{\rm iso}(\gamma)$
and the afterglow phase $E_{\rm iso}(t_{\nu_c =\nu_m})$ can be
determined and compared. For a two-sided jet, these isotropic values
are reduced by the factor $\theta_{jet}^2$/2. Thus, the
geometry-corrected gamma-ray energy $E(\gamma)=1.7\times 10^{51}$~erg
and the kinetic energy in the blastwave $E_k=3.2\times 10^{51}$~erg.
The value of $E(\gamma)$ differs from the compilation of Frail et
al.~(2001)\nocite{fks+01} because here we have used the circumburst
density derived from the broadband modeling rather than some assumed
value. Note also that $E_k$ is only a lower limit on the true initial
energy of the blastwave since $E_{\rm iso}(t_{\nu_c =\nu_m}$) is
derived at a time $t_{\rm \nu_{c} = \nu_{m}}$=1.4 days. After this
time the blast-wave evolution is predominantly adiabatic and the
energy dissipation is less than a factor of two up to 100 days after
the burst. We estimate that prior to this time (when radiative losses
decrease the blastwave energy) the energy drops by about a factor of
three. Another important quantity that can be estimated is
$\eta_\gamma$, the efficiency of the fireball in converting the energy
in the ejecta into $\gamma$ rays. A number of recent papers
(Beloborodov 2000\nocite{bel00}; Guetta, Spada \& Waxman
2001\nocite{gsw01}, Kobayashi \& Sari 2001\nocite{ks01}) have argued
that internal shocks under certain conditions are very efficient at
producing gamma-rays ({\em i.e.}, $\eta_\gamma\sim 0.2$).  From $E_k$
and $E(\gamma)$ we derive $\eta_\gamma\sim
E(\gamma)$/($E_k$+$E(\gamma)$) between 15\% and 35\%, comparable to
previous estimates of this and other well-studied events ({\em e.g.,}
Panaitescu \& Kumar 2001a\nocite{pk01b}).

\begin{figure}
        \epsscale{0.85}
  \plotone{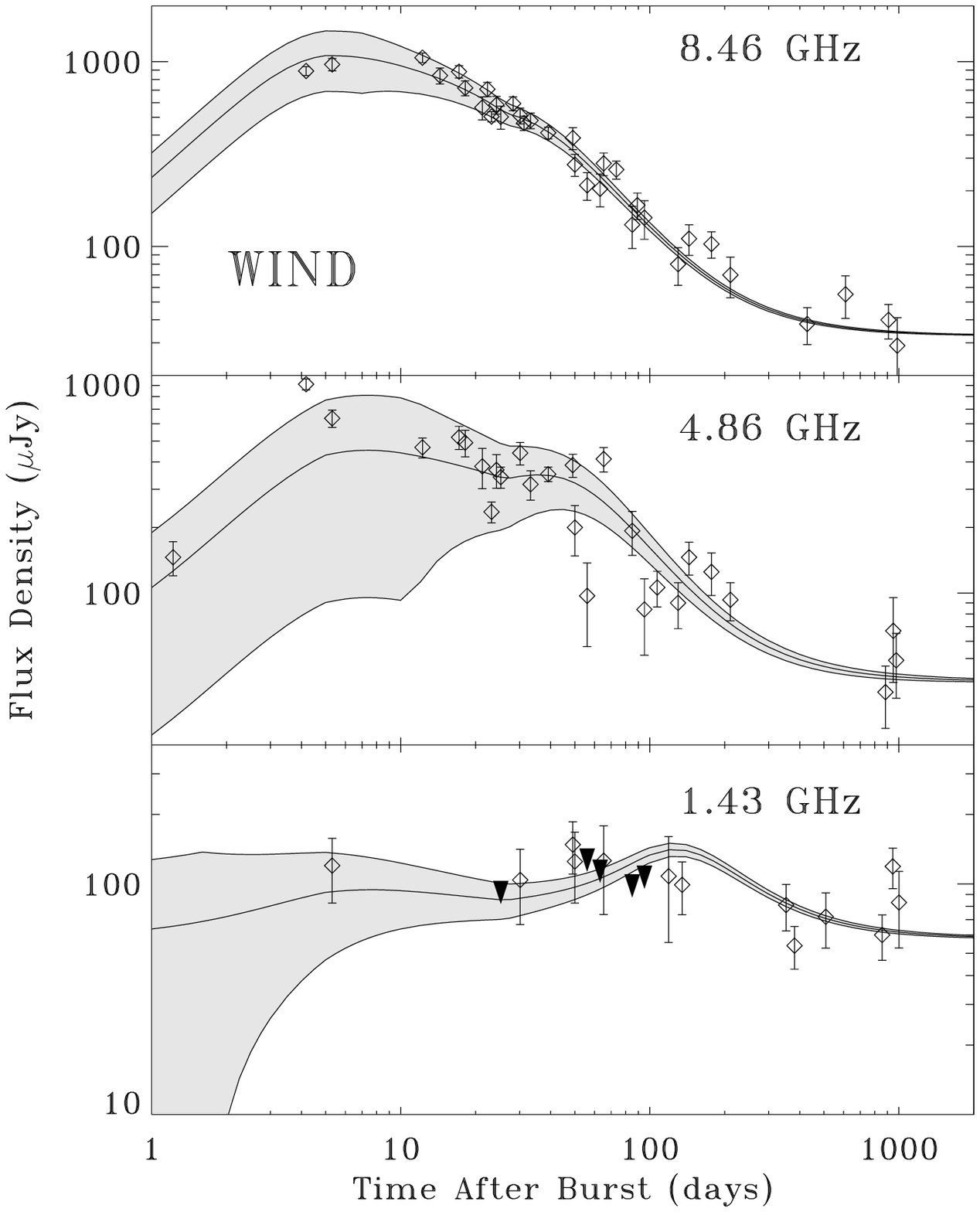}
\caption{Radio lightcurves of the \grb\ afterglow. The data is
  identical to that in Figure \ref{fig:radio} but the solid line is
  the WIND model discussed in the text.
\label{fig:wradio}}
\end{figure}

\begin{figure}
\plotone{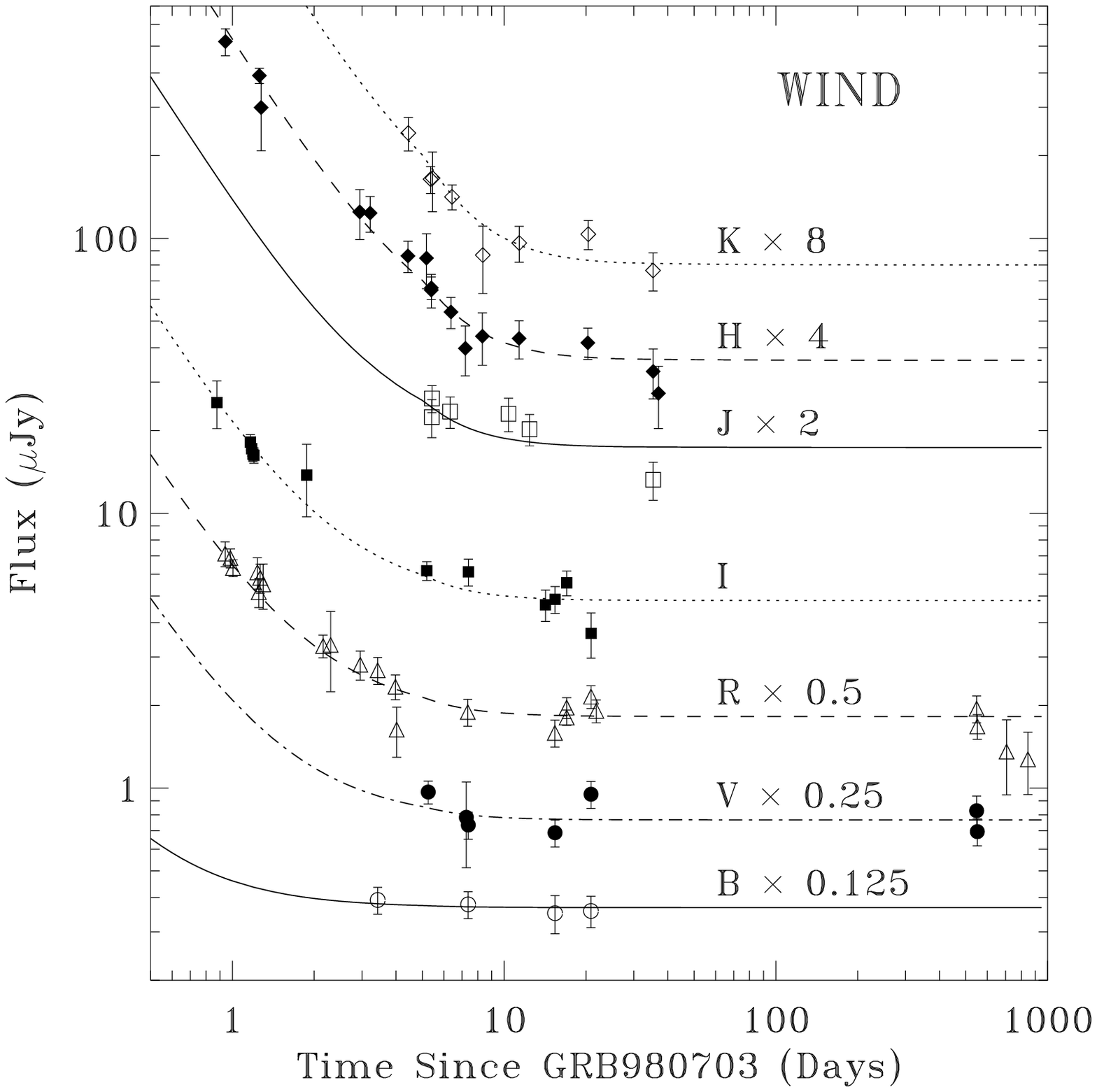}
\caption{Optical/NIR lightcurves of the \grb\ afterglow. 
  The data is identical to that in Figure \ref{fig:optical} but the
  solid line is the WIND model discussed in the text.
\label{fig:woptical}}
\end{figure}

\begin{figure}
  \plotone{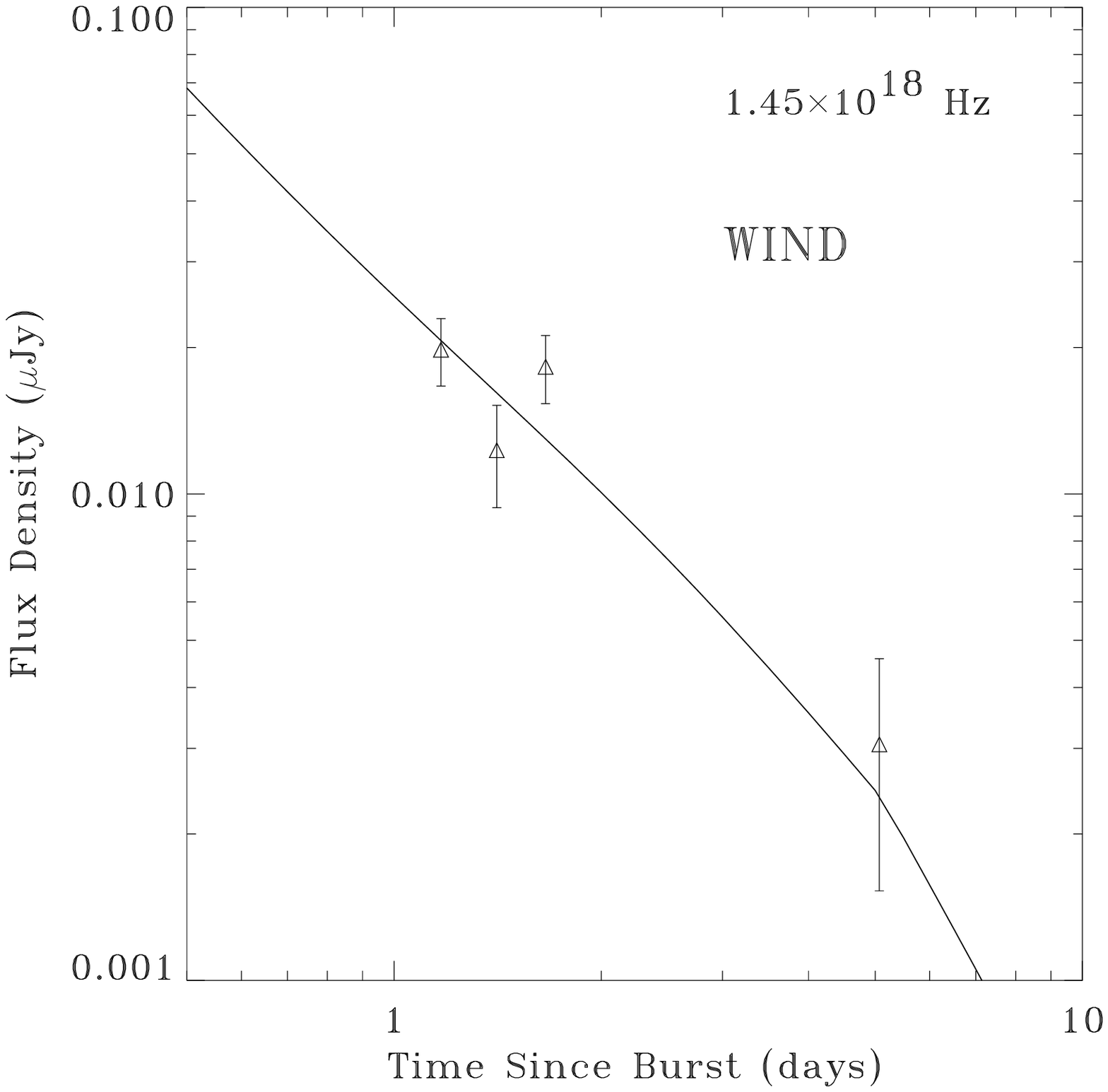}
\caption{The X-ray lightcurve of the \grb\ afterglow.
  The data is identical to that in Figure \ref{fig:xray} but the solid
  line is the WIND model discussed in the text.
\label{fig:wxray}}
\end{figure}

While this ISM model provides satisfactory agreement with the
broadband dataset (\S\ref{sec:bbdata}), it is not a unique solution.
An explosion into a wind-blown circumburst medium \cite{cl99} also
yields an equally good fit (see Table~\ref{tab:model} and
Figs.~\ref{fig:wradio}-\ref{fig:wxray}). The ejecta are also
collimated in this model with $\theta_{jet}\sim 18^\circ$. The density
is parameterized by $A_\star$ which characterizes the wind density,
with $\rho(R)=5\times 10^{11} A_\star R_{\rm cm}^{-2}$ g cm$^{-3}$,
with $R_{\rm cm}$ the wind radius in cm. The most troubling feature of
this model is that it requires about 70\% of the shock energy going
into the electrons.  Likewise, the geometry-corrected gamma-ray energy
of $E(\gamma)=3\times 10^{51}$~erg is a factor of 10 {\it larger} than
the kinetic energy in the blastwave $E_k$. This suggests an unusually
high $\eta_\gamma\simeq 90\%$, which, as noted above, is contrary to
theoretical expectations since little of the initial shock energy in
the fireball is left to power the afterglow. Thus, while a wind-blown
solution formally fits the data and cannot be ruled out, we prefer the
ISM model since it does not require such extreme physical conditions.

Regardless of which afterglow model is preferred, the host magnitudes
are comparable to those derived by Sokolov et
al.~(2001)\nocite{sfc+01} and Berger et al.~(2002)\nocite{bkf01} at
optical and radio wavelengths, respectively. Likewise, the steep
spectral slope $\beta_o$ (see \S\ref{sec:bbdata}) requires modest rest
frame V-band extinction A(V)$\sim$1, in accordance with earlier
estimates \cite{bfk+98,czg+99,vgo+99}.

\section{Comparison to Other Models}\label{sec:compare}

There have been several attempts to derive the fireball parameters for
\grb\ by constructing single-epoch spectra from the early afterglow
data (Bloom et al.~1998; Castro-Tirado et al.~1999; Vreeswijk et
al.~1999). The estimates for these parameters have varied widely among
these papers, due to slightly different data sets and a high degree of
correlation between the parameters. For example, there is a degeneracy
between the electron energy index $p$, the extinction A(V), and the
host brightness that makes it difficult to extract the underlying
spectral slope of the afterglow and therefore the location of two
important break frequencies $\nu_m$ and $\nu_c$.  This leads to large
uncertainties in the parameters E$_{iso}$, n, $p$, $\epsilon_e$ and
$\epsilon_B$.

The limitations of this spectral snapshot method can be overcome by
globally fitting all the afterglow data using a hydrodynamical model
of the blast wave. This is the approach that we have adopted in this
paper but the first application of this method to \grb\ was made by
Panaitescu \& Kumar (2001b)\nocite{pk01}. Their basic model is similar
to our own. They find that a collimated outflow in a constant density
medium provides a good description of the data, and they also find
acceptable fits to stellar wind model.  However, the differences
between our models show up most clearly in the derived fireball
parameters with E$_{iso}^{PK}=2.9\times 10^{54}$ erg,
n$^{PK}=7.8\times 10^{-4}$ cm$^{-3}$, $p^{PK}=3.08$,
$\epsilon_e^{PK}=0.075$, $\epsilon_B^{PK}=4.6\times 10^{-4}$, and
$\theta_{jet}^{PK}>0.047$ rad. Radiative losses are small in their
model and inverse Compton (IC) emission is negligible, while
$\epsilon_e$=0.27 in our model and IC is important for flattening the
X-ray light curve around day 1. The most severe difference, however,
is that the density derived by Panaitescu \& Kumar (2001b) is
3.5$\times 10^{4}$ times smaller than our estimate in
Table~\ref{tab:model}.

The origin of this discrepancy is not likely the result of differences
in the implementation of the relativistic blast wave model. Although
our specific methodology does differ somewhat, in at least one case
when fits were made using the {\it same} data for GRB\,000926, the
results were in good agreement \cite{hys+01, pk02}. The most serious
limitation of the Panaitescu \& Kumar (2001b) analysis of this burst
is that it relies on data taken over a limited frequency range and a
limited temporal range. The optical data were restricted effectively
to 1 to 5 days due to host galaxy contamination, and the early radio
data (especially at 5 GHz) were of limited use due to ISS. With the
addition of a complete set of centimeter radio light curves for \grb\ 
much of this difficulty can be resolved. The most significant area of
improvement is in the determination of the synchrotron self-absorption
frequency $\nu_a$. This important break frequency is largely
unconstrained in the Panaitescu \& Kumar (2001b) model and is likely
the origin of our discrepant density estimates.

\begin{figure}
  \plotone{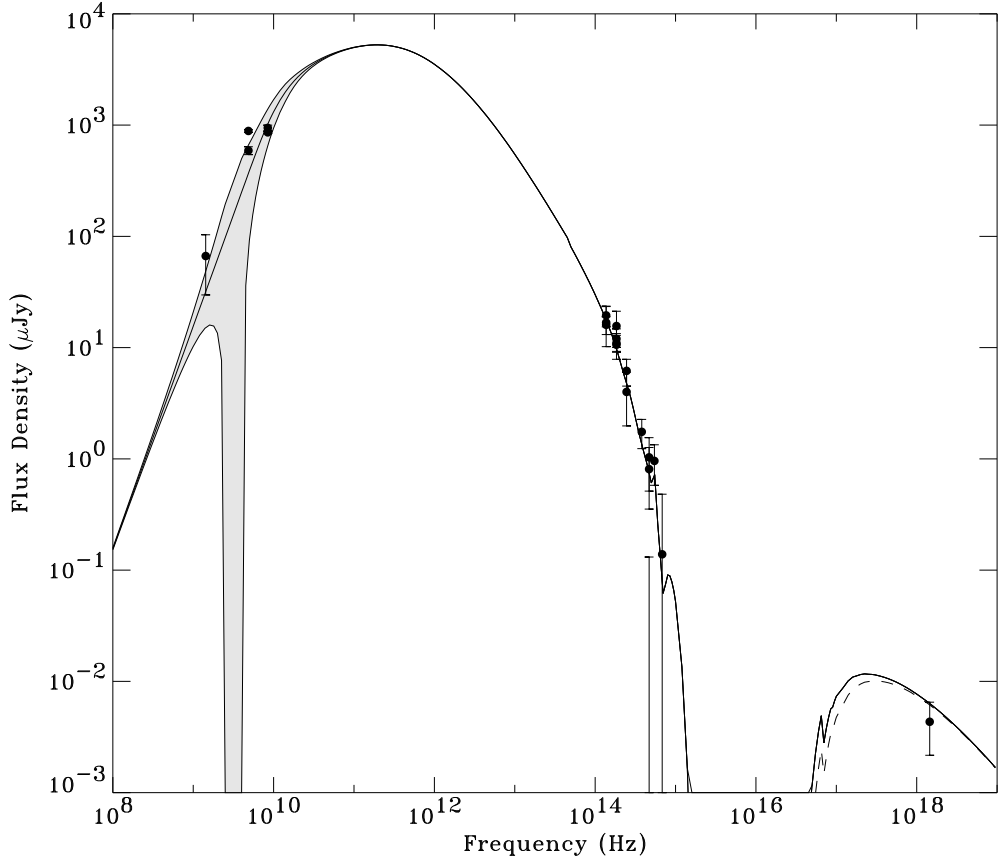}
\caption{The broadband spectrum of the \grb\ afterglow on day 4.5. All
  measurements taken between days 3.4 and 5.6 are included in the plot
  after scaling their flux densities to the epoch of 4.5 days using
  the ISM model. A host component has been subtracted from the
  optical/NIR and radio points (see Table~\ref{tab:model}). The solid
  line is the best-fit ISM model which includes a scintillation
  envelope in the radio band (shaded), host extinction in the optical
  band (A(V)=1.15), and a dominate inverse Compton component (dashed
  line) in the X-ray band.
\label{fig:spec}}
\end{figure}

An alternate way to view the difficulties in the afterglow model of
Panaitescu \& Kumar (2001b) is to use the ``C parameter'', introduced
by Sari \& Esin (2001)\nocite{se01}, which places a constraint on the
combination of synchrotron break frequencies and the peak flux
density.  From Fig. 1 of Panaitescu \& Kumar (2001b) we find the
following values for the synchrotron parameters: $\nu_m(t=1.2\,{\rm
  d})\approx 7\times 10^{12}$ Hz, $\nu_c(t=1.2\,{\rm d})\approx
3\times 10^{18}$ Hz, and $F_m(t=1.2\,{\rm d}) \approx 2$mJy. In order
not to violate the theoretial limit of $C<0.25$ it requires a
self-absorption break $\nu_a (t=1.2\,{\rm d})\ll 1$ GHz. It is this
upper limit on $\nu_a$ which leads to the low value of
n$^{PK}=7.8\times 10^{-4}$ cm$^{-3}$. A broadband spectrum of the
\grb\ afterglow on day 4.5 (see Fig. \ref{fig:spec}) shows this to be
a significant underestimate of $\nu_a$. If we use a more appropriate
value of $\nu_a$=14 GHz at this time than the additional synchrotron
parameters of Panaitescu \& Kumar (2001b) give an unphysical solution
with C$\gg$1 unless the cooling frequency $\nu_c$ is significantly
reduced and an IC component is added. This has the effect of
increasing the density of the circumburst medium.

\section{Discussion and Conclusions}

A high-quality panchromatic dataset, resulting from a multi-wavelength
observing campaign of \grb, has enabled us to apply the relativistic
blast wave model in order to determine the geometry and energetics of
the explosion, the density of the medium immediately surrounding the
progenitor, as well as the properties of the interstellar medium
within the host galaxy. All of the afterglow data for \grb\ are
consistent with a model in which the ejecta are collimated and
expanding into a constant density medium.  Although it is not a unique
solution, it yields reasonable estimates for the physical parameters
which are in agreement with other well-studied events.

Perhaps the most interesting result from this work is what has been
learned about the properties of GRB environment. A proper
understanding of the density structure of the circumburst medium
remains an important goal, since it is invariably tied to the GRB
progenitor question.  To the degree that the underlying assumptions
behind the fireball model of GRB afterglows are correct, broadband
modeling gives us the only {\it direct} determination of this density.
Optical extinction, host galaxy properties, X-ray lines, late-time
optical bumps, or the attenuation of low energy X-ray photons are all
{\it indirect} or line-of-sight measures of the GRB environment. In a
recent compilation of 10 well-studied afterglows, Panaitescu \& Kumar
(2002)\nocite{pk02} showed that broadband modeling yielded densities
in the range of 0.1-100 cm$^{-3}$. Their result is in good accord with
our own extensive modeling of afterglows ({\em e.g.,} Frail, Waxman \&
Kulkarni 2000\nocite{fwk00}, Berger et al.~2000\nocite{bsf+00}, Berger
et al.~2001\nocite{bdf+01}, Harrison et al.~2001\nocite{hys+01}, Yost
et al.~2002\nocite{yfh+02}).

For two events, GRB\ts{990123} and \grb, Panaitescu \& Kumar
(2001b)\nocite{pk01} the derived densities ({\em i.e.,} n$\lsim
10^{-3}$ cm$^{-3}$) are much lower than the values given above.  These
low estimates prompted the suggestion that some GRBs are massive stars
which explode in the pre-existing cavities of superbubbles created by
a previous generation of supernovae \citep{sw01}. The circumburst
density for \grb\ derived from our model, n$\simeq$28 cm$^{-3}$, is
much higher because the synchrotron self-absorption frequency
$\nu_{a}$ was not well constrained by the early observations.  On the
timescale of interest, $\nu_{a}$ lies within the radio band and is a
sensitive indicator of the ambient density, {\em i.e.,} $\nu_a\propto
n^{3/5} \epsilon_{e}^{-1} \epsilon_{B}^{1/5} E_{iso}^{1/5}$. A similar
problem likely explains results from GRB\ts{990123} but it is further
complicated by the evidence that the early radio emission was
dominated by a reverse shock component (Kulkarni et
al.~(1999)\nocite{kfs+99}. Likewise, it can be shown that the claims
of high circumburst densities ({\em i.e.,} n$>>10^{4}$ cm$^{-3}$)
based solely on X-ray and optical observations (Piro et
al.~2001\nocite{pgg+01}; in't Zand et al.~2001\nocite{iz+01}) cannot
be supported once radio data is included (Harrison et
al.~2001\nocite{hys+01}). Thus radio observations, which help to
constrain the low energy part of the synchrotron spectrum, are
essential for deriving accurate physical parameters of the blast wave.

In summary, for all well-studied GRB afterglows to date there is
little evidence for either extreme of high n$\gg10^{4}$ cm$^{-3}$ or
low n$\ll 10^{-3}$ cm$^{-3}$ circumburst densities. Instead, \grb\ is
the latest of a growing number of events whose density lies within a
narrow range of 0.1-100 cm$^{-3}$ with a canonical value of order
n$\sim$10 cm$^{-3}$. Such densities are found in diffuse interstellar
clouds of our Galaxy, commonly associated with star-forming regions. A
density of order 5-30 cm$^{-3}$ is also characteristic of the
interclump medium of molecular clouds, as inferred from observations
of supernova remnants in our Galaxy ({\em e.g.,} Chevalier
1999\nocite{chev99} and references therein).

\acknowledgements

DAF thanks thanks VLA observers Asantha Cooray, Rick Perley, Min Yun,
Harvey Liszt and Durga Bagri who made it possible to monitor this GRB
during the first few weeks. RS acknowledges support from the Fairchild
Foundation and from a NASA ATP grant. JSB acknowledges a grant from
the Hertz foundation. 


\newpage

\newpage
\begin{deluxetable}{lccccccc}
\tablecaption{Radio Flux Density History of \grb\tablenotemark{a}\label{tab:radio}}
\tablehead{ 
\colhead{Date} &
\colhead{$\Delta{t}$} & 
\colhead{F$_{8.46}$} & 
\colhead{$\sigma_{8.46}$} &
\colhead{F$_{4.86}$} & 
\colhead{$\sigma_{4.86}$} &
\colhead{F$_{1.43}$} & 
\colhead{$\sigma_{1.43}$} \\
\colhead{(UT)} & 
\colhead{(days)} & 
\colhead{($\mu$Jy)} &
\colhead{($\mu$Jy)} & 
\colhead{($\mu$Jy)} &
\colhead{($\mu$Jy)} & 
\colhead{($\mu$Jy)} &
\colhead{($\mu$Jy)} }
\tablecolumns{9} 
\startdata
1998 Jul. 04.40  &   1.22  &    \nodata & \nodata &  146 &  25 & \nodata & \nodata \\
1998 Jul. 07.35\tablenotemark{b}  &   4.17  &     890  & 21 &     912 &  26 &    \nodata &\nodata \\
1998 Jul. 08.49  &   5.31  &     965  & 55 &     635 &  49 &      120 & 37 \\
1998 Jul. 15.41  &  12.23  &    1050  & 35 &     467 &  43 &    \nodata & \nodata \\
1998 Jul. 17.56  &  14.38  &     840  & 72 &    1200 &  48 &    \nodata & \nodata \\
1998 Jul. 20.33  &  17.15  &     882  & 51 &     520 &  58 &    \nodata & \nodata \\
1998 Jul. 21.34  &  18.16  &     720  & 56 &     491 &  64 &    \nodata & \nodata \\
1998 Jul. 24.44  &  21.26  &     564  & 75 &     382 &  78 &    \nodata & \nodata \\
1998 Jul. 25.46  &  22.28  &     709  & 52 &   \nodata &  \nodata & \nodata & \nodata \\
1998 Jul. 26.33  &  23.15  &     504  & 20 &     236 &  23 &    \nodata & \nodata \\
1998 Jul. 27.42  &  24.24  &     584  & 57 &     368 &  62 &    \nodata & \nodata \\
1998 Jul. 28.41  &  25.23  &     502  & 67 &     341 &  34 &    25 & 39 \\
1998 Jul. 31.45  &  28.27  &     593  & 42 &   \nodata & \nodata &  \nodata & \nodata \\
1998 Aug. 02.30\tablenotemark{c}  &  30.12  &     580  & 60 &    \nodata &  \nodata & \nodata & \nodata \\
1998 Aug. 02.36  &  30.18  &     510  & 43 &     440 &  47 &      104 & 37 \\
1998 Aug. 03.44  &  31.26  &     465  & 34 &   \nodata & \nodata & \nodata & \nodata \\
1998 Aug. 05.41  &  33.23  &     480  & 40 &     316 &  46 &    \nodata & \nodata \\
1998 Aug. 11.29  &  39.11  &     412  & 22 &     352 &  21 &    \nodata & \nodata \\
1998 Aug. 21.28  &  49.10  &     386  & 49 &     387 &  43 &      148 & 37 \\
1998 Aug. 24.22  &  50.04  &     277  & 35 &     200 &  51 &      125 & 42 \\
1998 Aug. 28.22  &  56.04  &     214  & 35 &      97 &  40 &    68 & 37 \\
1998 Sep. 04.35  &  63.17  &     205  & 40 &   \nodata &  \nodata & 63& 32 \\
1998 Sep. 06.49  &  65.31  &     281  & 37 &     413 &  49 &  126 & 52 \\
1998 Sep. 14.57  &  73.39  &     260  & 26 &   \nodata & \nodata & \nodata & \nodata \\
1998 Sep. 26.22  &  85.04  &     131  & 33 &    193    &  43 &  18 & 46 \\
1998 Sep. 30.32  &  89.14  &     167  & 26 &   \nodata    &  \nodata & \nodata & \nodata \\
1998 Oct. 06.32  &  95.14  &     143  & 33 &    84     &  32 & 36 & 42 \\
1998 Oct. 18.31  & 107.13  & \nodata  & \nodata & 106  & 19 &\nodata & \nodata \\
1998 Oct. 30.24  & 119.06  & \nodata  & \nodata & \nodata  & \nodata & 108 & 52 \\
1998 Nov. 10.08  & 129.90  &      80  & 18 & 90 & 21 &\nodata & \nodata \\
1998 Nov. 15.03  & 134.85  & \nodata  & \nodata & \nodata  &\nodata & 99  & 25 \\
1998 Nov. 23.97  & 143.79  &     110  & 20 & 146 & 24 &  \nodata & \nodata \\
1998 Dec. 27.04  & 176.86  &     103  & 16 & 125 & 27 &  \nodata & \nodata \\
1999 Jan. 29.89  & 210.71  &      70  & 17 &  93 & 18 &  \nodata & \nodata \\
\enddata
\tablenotetext{a}{Each row lists the starting UT date of the
  observation, the time elapsed (in days) since the gamma-ray burst,
  the flux density (F$_\nu$) and the rms noise ($\sigma_\nu$) at
  8.46, 4.86 and 1.43 GHz.} 
\tablenotetext{b}{On 1998 July 7 we searched for polarized signal
  from the radio, obtaining 3-sigma limits on the linear and circular
  polarization at 4.86 GHz and 8.46 GHz of $\sim$8\%.}
\tablenotetext{c}{VLBA measurement. All other measurements made with
  the VLA. See \S\ref{sec:obs} for details.}
\end{deluxetable}
\normalsize

\newpage
\begin{deluxetable}{ccc}
\footnotesize
\tablecolumns{3}
\tablewidth{0pc}
\tablecaption{Model Parameters for \grb\label{tab:model}}
\tablehead{
\colhead{Parameter} & 
\colhead{ISM} & 
\colhead{Wind}}
\startdata
$\chi^2$ for 162 data pts & 170.4 & 171.4 \\
$t_{\rm jet}$ (days) & 3.43 & 5.11 \\
$t_{\rm non rel.}$ (days) & 49.6 & 26.4 \\
$t_{\rm \nu_{c} = \nu_{m}}$ (days) & 1.41 & 5.17 \\
$E_{\rm iso}(t_{\nu_c = \nu_m})(10^{52}$~erg)\tablenotemark{a} & 11.8 & 0.66 \\
n/A$^*$ & 27.6 & 1.42 \\
$p$  & 2.54 & 2.11 \\
$\epsilon_e$ (fraction of E) & 0.27 & 0.69 \\
$\epsilon_B$ (fraction of E) & 1.8$\times 10^{-3}$ & 2.8$\times 10^{-2}$ \\
$\theta_{jet} (rad)$ & 0.234 & 0.310 \\
\hline
host A(V) & 1.15 & 1.33 \\
host B ($\mu$Jy) & 2.93 & 2.94 \\
host V ($\mu$Jy) & 3.07 & 3.07 \\
host R ($\mu$Jy) & 3.61 & 3.64 \\
host I ($\mu$Jy) & 4.84 & 4.81 \\
host J ($\mu$Jy) & 8.77 & 8.67 \\
host H ($\mu$Jy) & 9.15 & 9.00 \\
host K ($\mu$Jy) & 10.1 & 10.0 \\
host 1.4 GHz ($\mu$Jy)  & 53 & 58 \\
\hline 
$E_{\rm iso}(\gamma)$\tablenotemark{b} (10$^{52}$~erg) & 6.01 & 6.01 \\
$E(\gamma)$ (10$^{50}$~erg) & 16.5 & 28.9 \\ 
\enddata
\tablenotetext{a}{Isotropic equivalent blastwave energy (not corrected for
collimation)}
\tablenotetext{b}{Isotropic-equivalent energy emitted in
the gamma-rays taken from Bloom, Sari \& Frail (2001)\nocite{bfs01}.}
\end{deluxetable}

\end{document}